\documentclass{aa}
\usepackage{graphicx}
\usepackage{txfonts}
\usepackage{natbib}
\newcommand{\taucet}{$\tau$\,Ceti } 
\newcommand{\epseri}{$\epsilon$\,Eridani } 
\newcommand{\tcet}{$\tau$~Cet } 
\newcommand{\tcets}{$\tau$~Cet}
\newcommand{\eeri}{$\epsilon$~Eri } 
\newcommand{\eeris}{$\epsilon$~Eri}
\newcommand{\micron}{\,$\mu$m}
\newcommand{\Kprime}{$K^{\prime}$}
\newcommand{\LD}{$\Theta_{\rm LD}$}

\newcommand{\teff}{$T_{\rm eff}$ }
\newcommand{\V}{$\cal V$}
\newcommand{\Mearth}{\,$M_{\oplus}$}
\newcommand{\Msun}{\,$M_{\odot}$}
\newcommand{\Rsol}{\,$R_{\odot}$}
\def\d{\mathrm d}
\begin{document}
   \title{A near-infrared interferometric survey of debris disk stars}

   \subtitle{I. Probing the hot dust content around \eeri and \tcet with CHARA/FLUOR}

   \author{
   E. Di Folco
          \inst{1}
          \and
          O. Absil\inst{2,6}
          \and
          J.-C. Augereau\inst{2}
          \and
          A. M\'erand\inst{3}
          \and
          V. Coud\'e du Foresto \inst{4}
          \and
          F. Th\'evenin\inst{5}
          \and
          D. Defr\`ere\inst{6}
          \and
          P. Kervella\inst{4}
          \and
          T.A. ten Brummelaar\inst{3}
          \and
          H.A. McAlister\inst{3}
          \and
          S.T. Ridgway\inst{7,3}
          \and
          J. Sturmann\inst{3}
          \and
          L. Sturmann\inst{3}
          \and
          N.H. Turner\inst{3}
          }

   \offprints{E. Di Folco}

   \institute{Observatoire de Gen\`eve, Universit\'e de Gen\`eve,
  Chemin des Maillettes 51, CH-1290 Sauverny, Switzerland\\
              \email{Emmanuel.Difolco@obs.unige.ch}
         \and
             Laboratoire d'Astrophysique de l'Observatoire de Grenoble, UMR CNRS/UJF 5571, BP 53, F-38041 Grenoble Cedex9, France
    \and
        Center for High Angular Resolution Astronomy, Georgia State University, PO Box 3969, Atlanta, Georgia 30302-3965, USA
    \and
             LESIA, UMR8109, Observatoire de Paris-Meudon, 5 place J. Janssen, F-92195 Meudon, France
    \and
    Laboratoire Cassiop\'ee, CNRS, Observatoire de la C\^ote d'Azur, BP 4229, F-06304 Nice Cedex 4, France
    \and
             Institut d'Astrophysique et de G\'eophysique, Universit\'e de Li\`ege, 17 All\'ee du Six Ao\^ut, B-4000 Li\`ege, Belgium
          \and
             National Optical Astronomical Observatory, 950 North Cherry Avenue, Tucson, AZ 85719, USA
             }

   \date{Received 10 April 2007; Accepted}


  \abstract
   {The quest for hot dust in the central region of
   debris disks requires high resolution and high dynamic
   range imaging. Near-infrared interferometry is a powerful means to directly
   detect faint emission from hot grains.
   }
   {We probed the first 3\,AU around \taucet  and \epseri
   with the CHARA array (Mt Wilson, USA) in order to gauge the
   $2$\micron~excess flux emanating from possible hot dust grains
   in the debris disks and to also resolve the stellar photospheres.
   }
   {High precision visibility amplitude measurements were performed with
   the FLUOR single mode fiber instrument and telescope pairs on baselines
   ranging from 22 to 241\,m of projected length. The short baseline observations
   allow us to disentangle the contribution of an extended structure from
   the photospheric emission, while the long baselines constrain the 
   stellar diameter.
   }
   {We have detected a resolved emission around \tcets, 
   corresponding to a spatially integrated, fractional excess flux of $0.98\pm0.21 \times 10^{-2}$
   with respect to the photospheric flux in the $K^{\prime}$--band. 
Around \eeris, our measurements can exclude a fractional excess of greater than $0.6\times 10^{-2}$~($3\sigma$). 
   We interpret
   the photometric excess around \tcet as a possible signature of hot grains in the inner debris disk
   and demonstrate that a faint, physical or background, companion can be safely excluded.
   In addition, we measured both stellar angular diameters with an unprecedented
   accuracy: $\Theta_{\rm LD}(\tau\,{\rm Cet})= 2.015 \pm 0.011$\,mas and
   $\Theta_{\rm LD}(\epsilon\,{\rm Eri})=2.126 \pm 0.014$\,mas.
   }
   {}


   \titlerunning{A near-infrared interferometric survey of debris disk stars. I}
   \maketitle
%

\section{Introduction}
Planetary systems around main-sequence (MS) stars were
first indirectly revealed through the detection of their far-infrared (IR)
photometric excesses by the IRAS and ISO satellites. Early interpreted
as the signpost of planetary activity, this IR emission is thought to arise
from short-lived dust grains in gas-free, optically thin disks. The production of grains
is sustained by asteroid collisions and out-gassing of comets in the first tens of AU
and by collisions of Kuiper belt-like bodies at larger distances \citep{mey07}.
Such grains are also known to form the zodiacal cloud in our own solar system.
Given the huge contrast with respect to their host star, so far only the brightest nearby
debris disks have been resolved. Single aperture images have highlighted
a great diversity of large scale structures comparable to our own Kuiper belt (KB)
 -- albeit more massive and extended-- with a relative void of matter in the
 central regions, where planets could plausibly have cleared away the dust
\citep[][and recent review by Mann et al.~2006]{sch06}. 

The detections of photometric excesses in the 25--100\micron~ range are much more numerous.
Recent surveys, that benefit from the high sensitivity of the Spitzer Space Telescope, have
revealed not only cold and distant grains in KB analogs, but also warmer grains in the inner disk
regions where planets could be present. For solar-type MS stars, the current photometric accuracy
of Spitzer places a $3\sigma$ detection threshold at 100 times the fractional luminosity $L_{\rm
dust} / L_{\star}$ of our zodiacal cloud at 70\,$\mu$m, 1000 times at 24\micron~ and 1400 times in
the $8-13$\micron~ range respectively. \citet{bry06} have reported a disk frequency of $13
\pm 5 \%$ at 70\micron, confirming the preliminary studies of \citet{dec00} and \citet{hab01},
with an extended and deeper survey of mature FGK field stars. At shorter wavelengths, \citet{bei06}
concluded that asteroid belts $10-30$~times more massive than our own are very rare, with an
8--13\micron~excess frequency lower than 2.5\,$\%$ for sun-like stars older than 1\,Gyr. This
result confirms the lack of warm ($T_{\rm gr} \gtrsim 300$~K) grains 
tentatively detected at mid-IR wavelengths by previous surveys.
Based on the current sensitivity limits, \citet{bry06} suggest that the statistical
distribution of positive detections is consistent with most nearby solar-like stars
harbouring exozodiacal clouds as bright as 0.1 to 10 times our own.

The presence of hot grains in the first AUs of extra-solar planetary systems cannot be
unequivocally determined by classical photometry. The typical accuracy of this method amounts to a
few percent at best in rather large fields of view (FOV). On the other hand, near-IR interferometers can 
provide sub-AU spatial resolution in a field of view comparable to the size of the telescope
diffraction pattern. Our observing strategy consists of using an optimised set of baseline
configurations in order to directly measure the excess ratio between the resolved disk and the only partially resolved
stellar photosphere \citep{edf04}. This method was recently shown to work with the
detection of a circumstellar \Kprime--band excess around Vega \citep{abs06}. Fiber-filtered
interferometers make it possible to directly measure contrasts larger than 100:1 in the near-IR.

The present paper investigates the case of the brightest sun-like stars in our neighborhood,
\epseri (K2V, 3.22\,pc, $K=1.7$) and \taucet (G8V, 3.65\,pc, $K=1.7$). These two stars have very
close spectral types but different ages: \eeri is younger than 1\,Gyr, while \tcet is about twice
as old as the Sun ($\sim10$\,Gyr) \citep{son00,hab01,edf04}. The cold regions of their debris disks
were imaged at sub-mm wavelengths. The images of \eeri reveal an almost face-on, clumpy
ring-like structure, peaking at 65\,AU \citep{gre98,gre05,pou06}. The disk around \tcet was
detected by \citet{gre04} at 850\micron~ and shows an elongated emission, that could be
interpreted either as an edge-on disk/ring extending out to 55\,AU, or as a face-on clumpy
structure. The inferred mass\footnote{assuming a collisional cascade of km-sized
bodies down to 20\micron~grains} amounts to 5--9\Mearth~around \eeri and 1.2\Mearth~around
$\tau$\,Cet, which can be compared with the 0.1\Mearth~content of cometary-like bodies in our Kuiper
belt \citep{gla01}.

We propose to further investigate this comparison with the inner solar system, by probing the warm
dust content of the interplanetary clouds around these two sun-like stars. This paper is
the first of a series aiming at directly detecting and characterising the emission of hot grains in exozodiacal
clouds with the CHARA Array interferometer \citep{teo05} on Mt~Wilson (California) with the
FLUOR beam combiner \citep[]{vcf97,mer06}. The survey that we have initiated focuses not only on solar-type stars, as described in the present paper, but also on earlier spectral types, as discussed in a forthcoming paper (Absil et al., in preparation).


\section{Interferometric observations and data analysis}
\subsection{CHARA / FLUOR observations}
\eeri and \tcet were observed in October 2006
in the $K^{\prime}$--band (1.94--2.34\micron).
Two pairs of 1\,m telescopes were used with separations of
34\,m (S1-S2 baseline) and 250\,m (S2-W1).
The short baseline allows the large scale emission around the stars (e.g., disk, companion)
to be resolved, while the long one constrains the stellar diameter through
the position of the first minimum of the visibility curve.
The FLUOR beam combining instrument measures the squared
modulus of the coherence factor (visibility) of the two interfering telescope beams,
after spatial filtering of the wavefronts by single-mode fibers.
The interferograms are recorded through a temporal modulation of the optical path
difference between the wavefronts with a 250\micron~scan length and a read-out
frequency of 500\,Hz. The instrumental visibility (or transfer function) was
estimated using interleaved observations of standard stars from the
catalogues of \citet{bor02} and \citet{mer05} (see Table~\ref{table_calib}). 
The interferometric field of view is limited to the Airy disk of individual telescopes,
i.e. about 0.55\,arcsec in radius at 2.2\micron~ (slightly chromatic across the \Kprime--band). 
The instrument field of view is further limited by the wider FOV of the single-mode fiber, 
which filters out the wavefront corrugations before the beam combination, 
and can thus be affected by a source of incoherent emission. 
The effective FOV results from the overlap integral of the turbulent wavefront 
with the Gaussian-shaped, fundamental mode of the fiber (FWHM~=~0.8\,arcsec). 
This results in a linear FOV of 2.6\,AU (FWHM) for \eeri and 2.9\,AU for \tcets.
  \begin{figure*}[t]
   \centering
   \includegraphics[width=8cm]{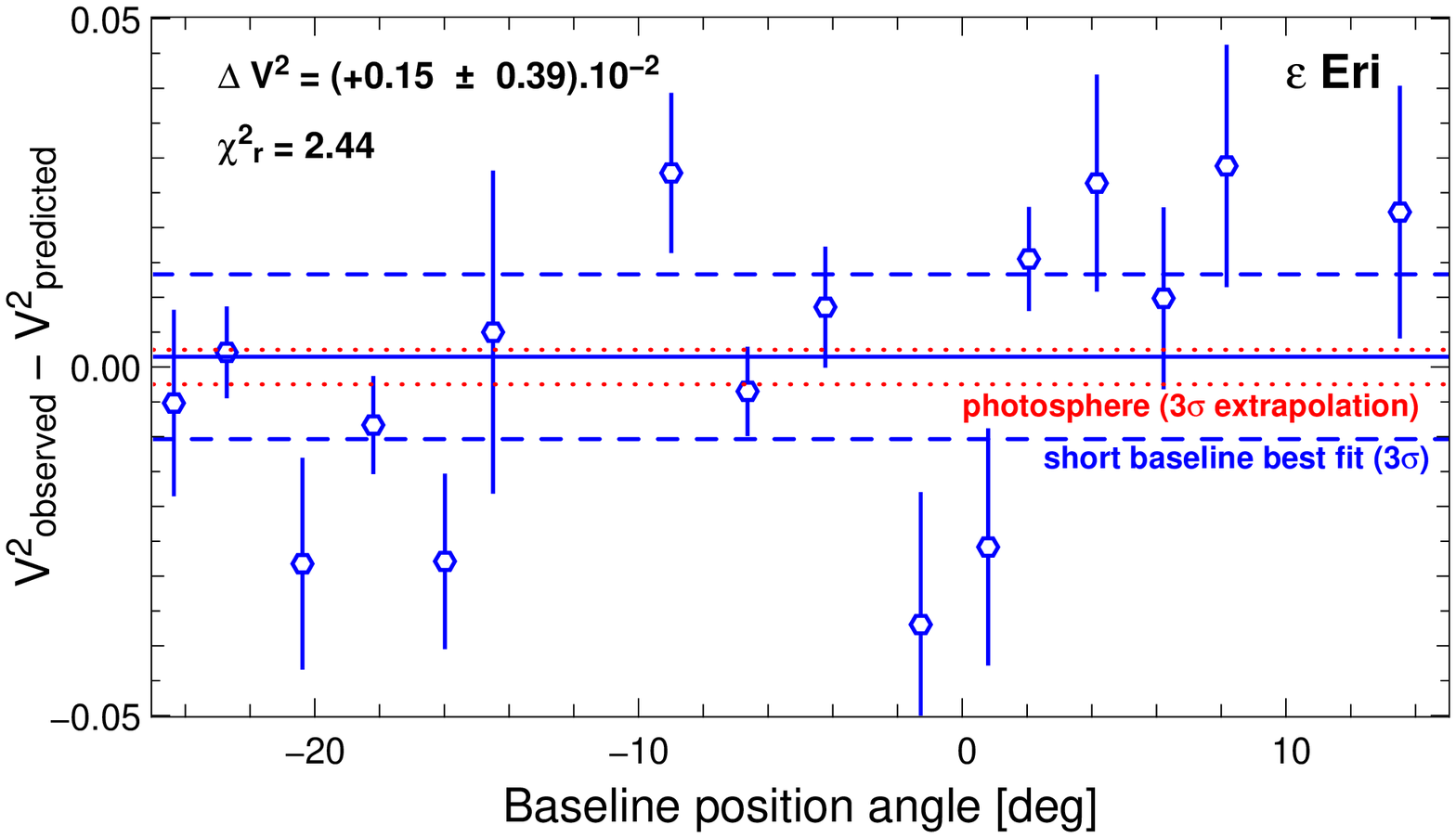}
\hspace{1cm}
   \includegraphics[width=8cm]{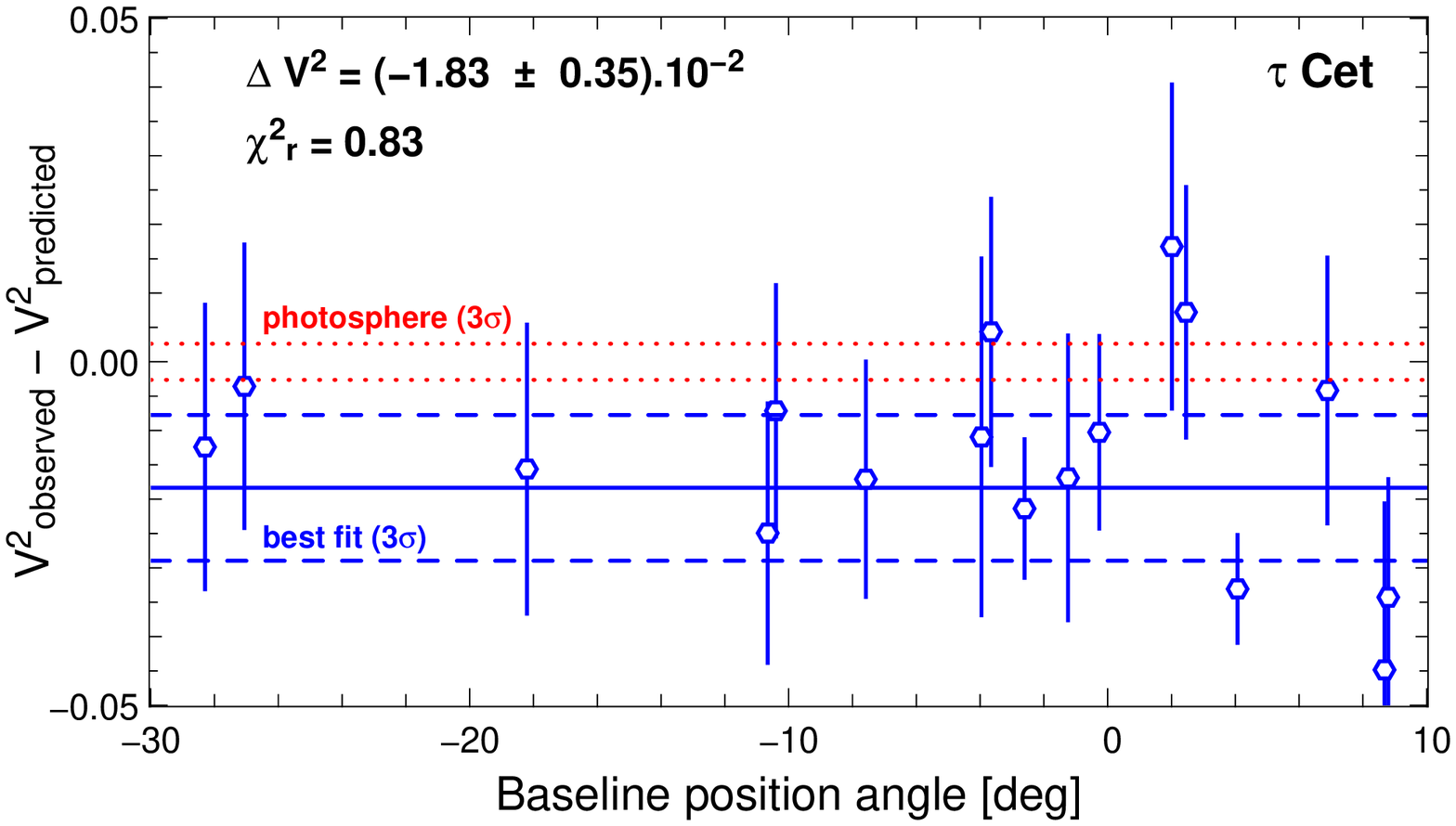}
      \caption{Squared visibility difference between FLUOR measurements (circles) on the 34\,m CHARA baseline and the predicted visibility of purely photospheric emission (based on VLTI/VINCI angular diameters). The thin dotted lines correspond to the visibility extrapolated from long baseline measurements and the solid lines represent the best fit of a constant \V$^2$ deficit to the new, short baseline data (with its $3\sigma$ interval -- dashed lines).}
         \label{Fig_deficit}
   \end{figure*}

\subsection{Data analysis and methodology}
\label{sec_analysis}
The broad-band squared visibilities (\V$^2$) are extracted using the FLUOR Data Reduction Software
\citep{vcf97,ker04,mer06} using a wavelet- and Fourier-based analysis.
The total uncertainty of each visibility measurement
includes the intrinsic dispersion of individual \V$^2$ estimations as well as the
calibrator induced error on the absolute visibility.

The visibilities of the source model are computed as follows:
\begin{equation}
{\cal V}^2(B)= \frac{\int{T_{\rm FLUOR}^2(\sigma)\space [{\cal B}(\sigma, T_{\rm eff})/\sigma ]^2\space {\cal V}^2_{\sigma}(B)\space d\sigma}}{\int{T_{\rm FLUOR}^2(\sigma) \space [{\cal B}(\sigma, T_{\rm eff})/\sigma ]^2 \space d\sigma}}, \quad \sigma = 1/ \lambda
\label{eqn_WB}
\end{equation}
where $T_{\rm FLUOR}(\sigma)$ is the FLUOR chromatic transmission
function in the \Kprime--band filter, ${\cal B}(\sigma, T_{\rm eff})/\sigma$
is the Planck function (in photons s$^{-1}$ Hz$^{-1}$ m$^{-2}$)
for the stellar effective temperature $T_{\rm eff}$ and ${\cal V}^2_{\sigma} (B)$
is the monochromatic visibility of the source at the projected baseline $B$. 
This formula takes into account the so-called bandwidth smearing effect, 
the result of which is that the modeled wide-band visibilities never reach zero.

While compact photospheres with a typical angular diameter of $\theta_{\star} \sim 2$\,mas are
fully resolved with baselines of order of 250\,m at $\lambda = 2$\micron, an extended emission from
possible hot grains beyond their sublimation distance, over resolved at these large baselines, 
can be resolved with telescope separations
as short as 10--20\,m. In order to detect such an IR excess, we first compare FLUOR visibilities at
the 34\,m baseline with the visibility expected for a purely photospheric emission. To this
end, we use the high precision angular diameters derived from earlier interferometric observations
with VLTI/VINCI \citep{edf04} to extrapolate the visibility of the stellar photosphere for the 20--30\,m range.
At such low spatial frequencies, the extrapolated \V$^2$ is 
insensitive to the photospheric limb-darkening profile, so the uniform disk approximation can be used.
Since the stellar surface is almost completely unresolved (${\cal V}^2_{\star} \gtrsim 0.9$), this 
extrapolation can be done with high precision. 

In a second step, a simple model of the source brightness distribution is fitted to the full data
set, including short and long baselines, in order to simultaneously estimate the photospheric angular
diameter and the contrast of the circumstellar environment (CSE), if any.
The photospheric limb-darkening follows the, non physical, description
proposed by \citet{hes97}: $I_{\lambda} = \mu^{\alpha}$, were $\mu$ is
the cosine of the azimuth of a surface element of the star where
$\mu =1$ at the centre of the stellar disk and 0 at the limb.
This parameterization results in an analytical formulation for the
monochromatic stellar visibility:

\begin{equation}
{\cal V}_{\sigma, \star} (x) = \Gamma(\nu+1) \space \frac{J_{\nu}(x)}{(x/2)^{\nu}}\mathrm{\, , with\,\,} x= \pi \theta_{\star} B \sigma \mathrm{\,\,and\,\,} \nu =\frac{\alpha}{2} +1
\end{equation}

If a significant deficit of visibility is detected at low spatial frequencies, the global model
takes into account a second component in addition to the star itself. Writing $f_{\rm CSE}$
the relative brightness of the CSE with respect to the total brightness, the final squared
visibility can be written as:

\begin{equation}
{\cal V}^2 (B) = \big[ f_{\rm CSE} {\cal V}_{\rm CSE}(B) + (1-f_{\rm CSE}) {\cal V}_{\star}(B) \big]^2
\end{equation}

If we assume the CSE to be composed of a uniform emission of infinite radius
(or at least extended enough to be over resolved at the shortest baseline, i.e. 
$V_{CSE} (B) \simeq 0 $ for all $B > B_{\rm min}$), this expression can be simplified to:

\begin{equation}
{\cal V}^2 (B) \simeq (1- f_{\rm CSE})^2 {\cal V}^2_{\star}(B) \qquad (B > B_{\rm min}),
\label{eqn_disk}
\end{equation}
where ${\cal V}^2_{\star}$ is the broad-band visibility computed as in~(\ref{eqn_WB}). In
particular,
${\cal V}^2 (B) \simeq (1- 2 f_{\rm CSE}) {\cal V}^2_{\star}(B)$ if $f_{\rm CSE} \ll 1$, and the 
visibility deficit $\Delta {\cal V}^2$ at short baselines is about twice as large as the CSE
brightness ratio.

For the sake of simplicity, and given our limited sampling in spatial frequencies, we will assume
the source brightness profiles to be circularly symmetric. For \eeris, this is supported by the
independent estimations of the inclination of its stellar rotation axis \citep[$30 \pm
3\deg$,][]{cro06} and of its debris disk \citep[about 25\,$\deg$,][]{gre05}. The inclination of
\tcet is less clear, but both stars have very low rotational velocities, with $v_{\rm rot}
\sin{i}$ respectively equal to 1.4\,km.s$^{-1}$ for \tcet and 2.4\,km.s$^{-1}$ for \eeri
\citep{val05}, so that no detectable elongation of their stellar surface is expected.

\subsection{Results for \taucet and \epseri}
\label{sec_results} The results of the short baseline measurements are presented in
Fig.~\ref{Fig_deficit}. The projected baseline ranges between 22 and 26\,m for \tcet and between 24
and 28\,m for \eeris. These plots show a clear departure from the expected photospheric visibility for \tcets. The visibility extrapolated from the earlier estimation of the stellar diameter with
VLTI/VINCI is plotted as a thin dotted line for reference ($3\sigma$ interval).
We fit a \V$^2$ deficit ($\Delta {\cal V}^2$\,$={\cal V}^2_{\rm measured} -{\cal V}^2_{\rm predicted})$, 
which is assumed to be constant over the projected baseline range: $\Delta {\cal
V}^2\,$(\tcets)$=(-1.83 \pm 0.35) \times 10^{-2}$ ($\chi^2_{\rm r} =$0.83). The data do not show
any trend associated with the baseline azimuth, which supports the assumption of circular symmetry.
In the case of \eeris, the measurements are consistent with
the visibility expected from the VLTI/VINCI estimation of the stellar angular diameter. Although
the dispersion of the measurements is slightly larger than in the case of \tcets, we can put a
$3\sigma$ upper limit to the visibility deficit in the 20--30\,m range $|\Delta {\cal
V}^2$($\epsilon$\,Eri)$|<1.17 \times 10^{-2}$ ($\chi^2_{\rm r} =2.4$). 
The rather high $\chi^2_{\rm r}$ value is due to the dispersion of the measurements, 
with a large contribution from the five points with larger uncertainties 
but beyond $2.5\sigma$ from the central fit value. The larger dispersion 
compared to \tcet\ seems to be linked to a degraded coupling of light into the instrument fibers, 
possibly due to poorer atmospheric conditions.
\begin{figure*}[t]
   \centering
   \includegraphics[width=8cm]{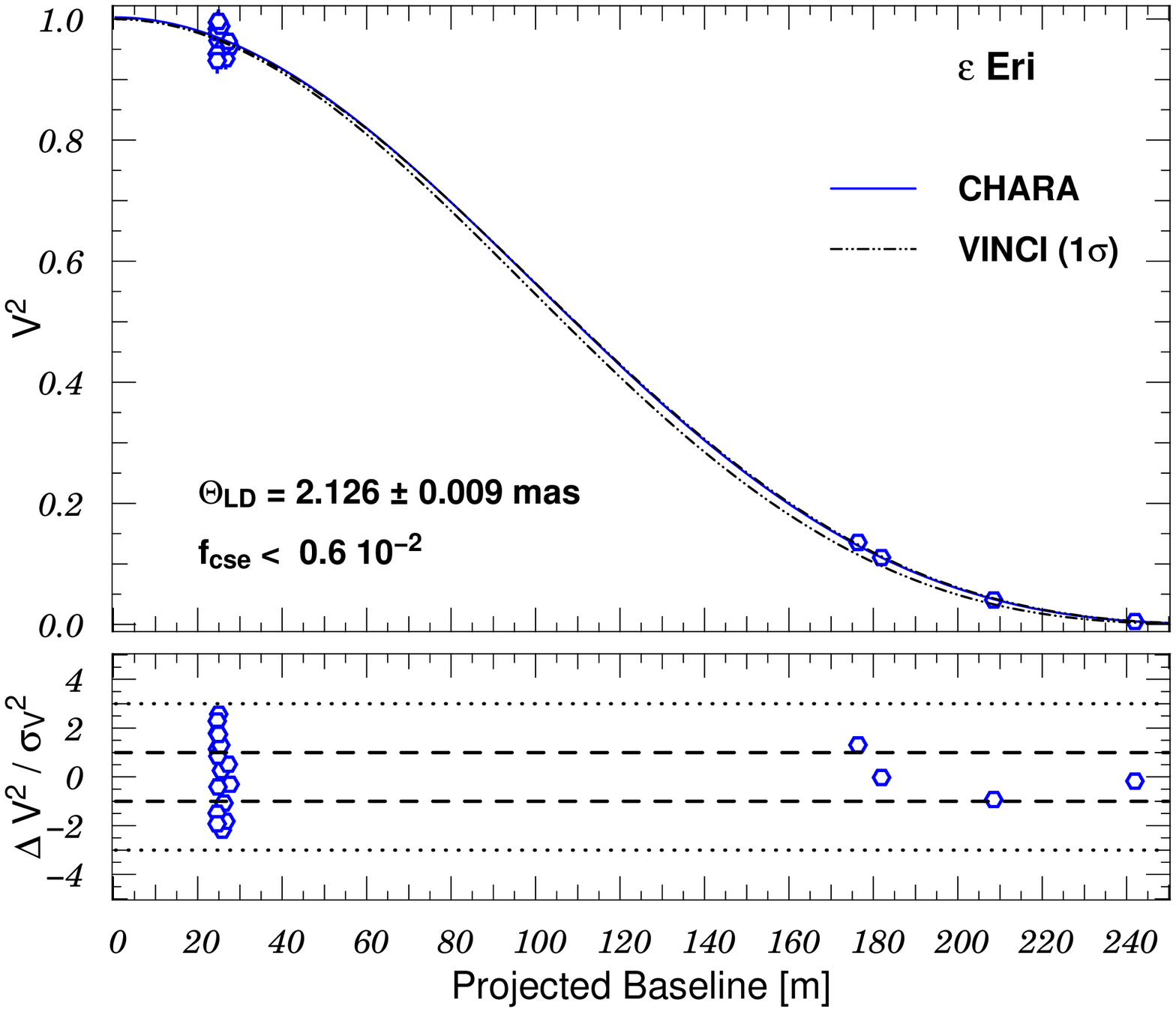}
   \hspace{1.25cm}
   \includegraphics[width=8cm]{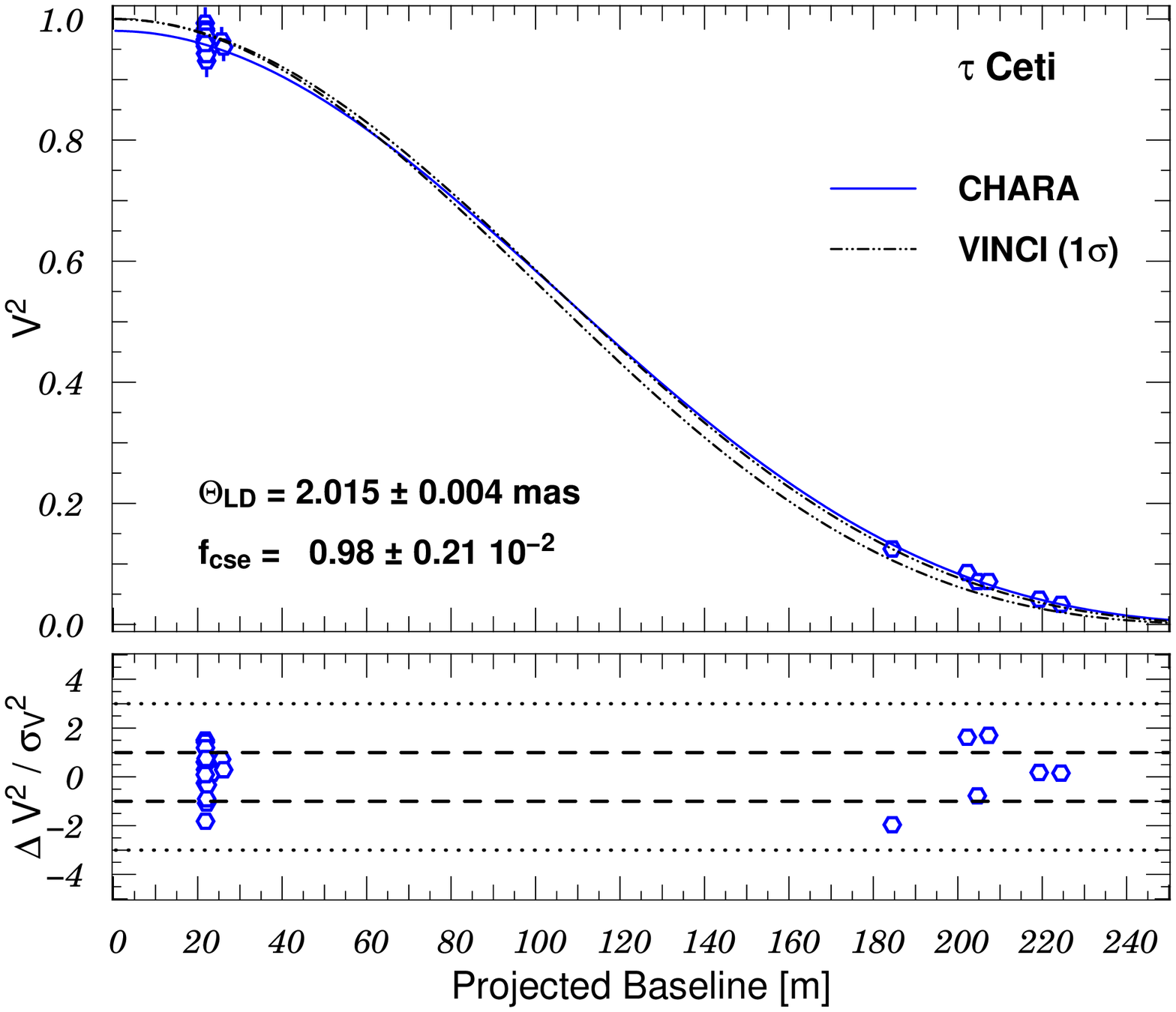}
       \caption{Observations of \epseri (left) and  \taucet (right) with FLUOR on CHARA baselines (S1-S2 and S2-W1). The solid line is the best fit of a source model consisting of a limb-darkened photosphere plus a fully resolved disk. The ($1\sigma$) visibility expected from the earlier VINCI measurements is represented by the dot-dashed line. The errors on the LD diameters are the $1\sigma$ statistical uncertainties from the fit. The lower part of the panels presents the distance between each single measurement and the best fit (dashes stand for the $1\sigma$ limit and dots for the $3\sigma$)}.
         \label{Fig_V2}
   \end{figure*}
\begin{table}[b]
\caption[]{Relevant parameters for the calibrator stars. \label{table_calib}}
\begin{tabular}{lcccccl}
\hline
Identifier & Sp. type & $m_{\mathrm{K}}$ & \LD~$\pm 1 \sigma$ [mas] & Target \\
\hline
\noalign{\smallskip}
HD 787 & K5III &  1.80 & 2.52 $\pm$ 0.03 $^{\mathrm{a}}$ & \tcet\\
HD 1522 & K1III & 0.92 & 3.36 $\pm$ 0.04 $^{\mathrm{a}}$ & \tcet\\
HD 6805 & K2III & 0.88 & 3.44 $\pm$ 0.04 $^{\mathrm{a}}$ & \tcet\\
HD 8815 & K0IIIb & 1.48 & 2.76 $\pm$ 0.03 $^{\mathrm{a}}$ & \tcet, \eeri\\
HD 12596 & K2III & 3.12 & 1.233 $\pm$ 0.016 $^{\mathrm{b}}$ & \tcet\\
HD 12685 & M0/1III & 2.83 & 1.450 $\pm$ 0.019 $^{\mathrm{b}}$ & \tcet\\
HD 16212 & M0III & 1.28 & 3.11 $\pm$ 0.03 $^{\mathrm{b}}$ & \tcet, \eeri\\
HD 16526 & K4/5III & 2.92 & 1.389 $\pm$ 0.017 $^{\mathrm{b}}$ & \tcet\\
HD 18071 & G8/K0III & 3.09 & 1.056 $\pm$ 0.014 $^{\mathrm{b}}$ & \tcets, \eeri\\
HD 29063 & K2III & 2.80 & 1.38 $\pm$ 0.017 $^{\mathrm{b}}$ & \eeri\\
HD 39853 & K5III & 1.84 & 2.38 $\pm$ 0.027 $^{\mathrm{a}}$ & \eeri\\
HD 220935 & K5III & 2.6 &  1.963 $\pm$ 0.098 $^{\mathrm{b}}$ & \tcet\\
HD 221745 & K4III & 2.88 & 1.426 $\pm$ 0.019 $^{\mathrm{b}}$ & \tcet\\
\hline
\end{tabular}
\begin{list}{}{}
\item[$^{\mathrm{a}}$] From \citet{bor02}.
\item[$^{\mathrm{b}}$] From \citet{mer05}.
\end{list}
\end{table}

The combined data collected at short and long baselines (up to 224\,m for
\tcet and 241\,m for \eeris) are then fitted with a limb-darkening model of
photospheric emission, along with a circumstellar component in the case of \tcets,
as explained in details in Sect.~\ref{sec_analysis}. The $\alpha$ values for the
limb-darkening (LD) description of Hestroffer are derived from the theoretical 4-parameter
law tabulated by \citet{cla00} for the adopted stellar \teff and metallicity.
We find $\alpha=0.15$ for \tcet and $\alpha=0.16$ for \eeris.
Because of our sparse sampling of the ($u,v$) plane, our data do not allow us to spatially
characterise the shape of the detected CSE around \tcets. We therefore consider the
simple case of a uniformly bright component in the FOV (i.e. a fully resolved
disk), accounting for a fractional -- spatially integrated -- flux ratio $f_{\rm CSE}$.
The model visibility, including the bandwidth smearing effect, then follows from
Eq.~\ref{eqn_disk}.

For \eeris, the data are consistent with no excess emission from the CSE, within a $0.2\times 10^{-2}$
uncertainty. The LD model fit results in an angular diameter of \LD~$= 2.126 \pm 0.009$\,mas 
(statistical error only). The fractional excess emission of its inner debris disk can be
constrained to values lower than $0.6\times 10^{-2}$ ($3\sigma$ upper limit). We emphasize that the presence
of a faint excess due to the CSE not only produces a visibility deficit at short baselines, but
also changes the general slope of the visibilities at long projected baselines. Therefore
it can bias any angular diameter estimation by an amount that exceeds the internal
dispersion of the measurement \citep[for more details, please refer to][]{edf04}. In the case of
\tcets, the relative CSE excess amounts to $f_{\rm CSE}= 0.98 \pm 0.21\times 10^{-2}$ and the
de-biased CHARA LD diameter is estimated to be \LD~$= 2.015 \pm 0.004$\,mas. Correcting the 2004 VLTI
data for this bias leads us to the revised value \LD~$= 2.059 \pm 0.033$\,mas. In the next section, we analyse the error
budget and further compare the CHARA and VLTI measurements.

\subsection{Discussion}
\eeri and \tcet are very similar in spectral type, magnitude and declination, and they
also share several calibrator stars, so the potential instrumental or calibration bias in our
measurements should be the same for both targets. The detection of a short baseline deficit for
\tcets, associated with a negative detection for \eeris, and a comparable statistical uncertainty,
reinforces our confidence in the observing method and in the reality of the detection itself.
The observed CSE excess is of the order of the excess detected around the A0 dwarf Vega 
\citep[$f_{\rm CSE}=1.29 \pm 0.19\times 10^{-2}$,][]{abs06} and allows the stellar angular diameter to be debiased. The derived
uncertainties confirm that we can reach a precision on the CSE excess of about $0.2 \times 10^{-2}$ 
($1\sigma$) in 10--15 measurements, under median atmospheric conditions for $K=2$ stars.

The formal statistical precision
of the fitted angular diameters is very small (5--10\,$\mu$as), since we were able
to approach the first minimum of the visibility function. The total uncertainty
should also include contributions from the uncertainty of the
calibrator diameters, the uncertainty of the chromatic transmission function
$T_{\rm FLUOR}$, and from our a priori assumptions on the limb-darkening profiles.
For the calibration uncertainty we use a conservative value of 0.010\,mas.
The uncertainty of the stellar metallicity (0.2\,dex) and \teff (100\,K) leads to a typical
uncertainty for the $\alpha$-determination of $5 \times10^{-3}$, which translates into a maximum
2.5\,$\mu$as error on the fitted angular diameters. The LD parameters tabulated by \citet{cla00}
are based on ATLAS 1D-models of stellar atmospheres. \citet{big06} have compared the determination of diameter using
the 1D ATLAS model to those produced by the 3D radiative hydrodynamical simulations for the star
$\alpha$\,Cen\,B, a star whose fundamental parameters are very close to those of our targets (K1V,
\teff=$5260 \pm 50$\,K, $\log g =4.5$, [Fe/H]=0.2). They find that the 3D simulations produce a
systematically smaller diameter (17\,$\mu$as difference, or 0.3\,\%), mainly impacting the second
lobe visibilities. Finally, for the uncertainty of the instrumental transmission, 
we use the error quoted by \citet{big06} for the VINCI instrument, a conceptual copy of the 
FLUOR instrument developed for the VLTI. This 0.15\,\% uncertainty
leads to an additional 3\,$\mu$as error on our angular diameters. Quadratically adding the
statistical precision from the fits and the systematics related to the calibration, to the
$\alpha$-prescription and to the chromatic transmission, we derive a total uncertainty of
0.011\,mas for \tcet and 0.014\,mas for \eeris. We emphasise that the use of the 1D ATLAS models
for the a priori assumption on the LD profile, which are not constrained at all by our observations, may
overestimate these diameters by an amount comparable to the final uncertainty, compared to 3D
hydrodynamical models. Measurements in the second lobe of the visibility function will be needed to
properly estimate the impact of the LD profile on the diameter determination.
\begin{table}
\caption[]{Global model fits and comparison with earlier estimations. \label{table_results}}
\begin{tabular}{lcccccl}
\hline
Target & Instrument & \LD~$ \pm \sigma$ [mas] & $f_{\rm CSE} \pm \sigma$ ($10^{-2}$) & $\chi^2_{\rm r}$ \\
\hline
\noalign{\smallskip}
\tcet & CHARA & $2.015 \pm 0.011$  & $0.98 \pm 0.21$ & 1.1 \\
 & VLTI$^{\mathrm{a}}$ & $2.078 \pm 0.031$  & $< 5$ \\
 & VLTI$^{\mathrm{b}}$ & $2.059 \pm 0.033$  & 0.98 \\
\noalign{\smallskip}
\hline
\noalign{\smallskip}
\eeri & CHARA & $2.126 \pm 0.014$ & $< 0.6$ & 2.1 \\
 & VLTI$^{\mathrm{a}}$ & $2.148 \pm 0.029$ & $< 2$ \\
\hline
\end{tabular}
\begin{list}{}{}
\item[$^{\mathrm{a}}$] From \citet{edf04}.
\item[$^{\mathrm{b}}$] New fit with fixed $f_{\rm CSE}$ value.
\end{list}
\end{table}

A comparison between CHARA/FLUOR and VLTI/VINCI measurements, taking into account the total error budget, is presented in
Table~\ref{table_results}. After a correction for the
CSE bias, VINCI diameter estimations appear to agree with the new values within
$1.3\sigma_{\rm vinci}$ for \tcet and within $0.8\sigma_{\rm vinci}$ for \eeris. The
precision of the Hipparcos parallax contributes 12\,\% to the final uncertainty on the
linear radius of \tcet and 8\,\% for \eeris: $R(\tau \, {\rm Cet}) = 0.790\pm 0.005$\,\Rsol~and 
$R(\epsilon \, {\rm Eri}) = 0.735 \pm 0.005$\,\Rsol.
\section{Nature of the infrared excess}
\subsection{An unlikely faint companion around \tcet}
\label{sec_binary}
Several physical models can explain the visibility deficit observed at short baselines.
One trivial solution can be an IR point source within the
FOV of the instrument. Deriving the contribution of this source
from the amplitude of the visibility deficit allows us to address
the question of its angular separation with the scientific target.
A maximum upper limit on the source contrast
can be set to $f \sim \Delta$~\V$^2 /4 = (0.46 \pm 0.09)\times 10^{-2}$,
equivalent to a maximum observed magnitude $m_{K_2}=7.6 \pm 0.3$.

For a physically bound faint source, we can estimate the spectral type from the known distance of
\tcet ($3.65 \pm 0.01$\,pc). Applying the empirical relations from \citet{del00} with borderline conditions,
it follows that a companion would have an absolute magnitude of
$M_K = 9.7 \pm 0.3$, and thus a mass $M \simeq 0.09 \pm 0.01$\,$M_{\odot}$. 
Using an age of about 10\,Gyr and a metallicity $\rm{[M/H]} = -0.5$, \citet{bar98} models indicate a temperature
\teff~=~2800--3000\,K and a spectral type M6-9V. Orbiting at less than 4\,AU, such a massive body
would also present a clear astrometric and/or radial velocity (RV) signature, depending on the
inclination of the system. With a mass ratio of 10 and a maximum separation of 4\,AU (period$<
7.6$\,yr), the astrometric signature of the binary would amount to $\alpha \geq 110$\,mas (minimum
angular semi-major axis). No significant motion was detected by Hipparcos at a 0.8\,mas level
during the 4-year mission \citep{per97}. Moreover, long-term radial velocity follow-up of \tcet 
has reported very stable measurements over about a decade. \citet{wit06} exclude the presence of a
Jupiter-mass planet on a 5\,AU, circular orbit based on an upper limit of the semi-amplitude
velocity $K_{\rm RV}\sim$10--15\,m.s$^{-1}$. This is also independently confirmed by the CORALIE
survey which shows a constant radial velocity with a 5\,m.s$^{-1}$\,rms over more than 5\,yr (Udry,
private communication). RV measurements alone would constrain the orbital inclination (with respect to  the sky plane)
of any 0.09\Msun~companion to values smaller than 0.21\,deg, which is very unlikely and
also incompatible with the astrometric constraint.

Furthermore, an IR background source can also be ruled out given
its low statistical likelihood. The 2MASS survey \citep{cut03}
has detected only 355 sources brighter than $K=8$ in a 5\,deg radius
patch around $\tau$\,cet. The local surface density of such IR sources
is thus as low as $3 \times 10^{-7}$\,arcsec$^{-2}$, hence the probability of 
finding such a faint source in a maximum FOV of about 2\,arcsec$^2$ is $6 \times 10^{-7}$.
The same reasoning leads to a probability of $9 \times 10^{-7}$ for \eeris. 
In conclusion,
although our interferometric observations alone could in principle
be reproduced with an additional faint point-like source in the FOV,
we can confidently exclude the presence of any such companion.
\subsection{Circumstellar hot dust grains around \tcet}
\label{sec_disk_simu} We show in this section that the detected \Kprime--band excess around \tcet
can be explained by the emission within our FOV of hot dust grains in an exozodiacal disk.
Since the spatial frequencies probed by the current data set do not allow us to determine the exact geometry of the
emission, we assume it to be represented by a disk of outer radius as large as our
FOV, or at least large enough to be fully resolved at $B\sim20$\,m. Following
Eq.~\ref{eqn_disk}, the derived integrated flux excess of this inner disk, relative to the
stellar flux in the \Kprime--band, is $f_{\rm CSE} = 0.98 \pm 0.21\times 10^{-2}$. Other models such as a
thin ring or a spherical envelope could also be considered and would provide slightly different
values for $f_{\rm CSE}$, yet have a small impact on our final result. However, the spatial
characterisation of this CSE is beyond the scope of this paper, since it would require many
more visibility measurements with a continuous sampling of the ($u$, $v$) plane. Based on this
photometric excess value, we will use a modeling approach similar to that described
in \citet{abs06}, and successfully used to reproduce the resolved emission around Vega.

\begin{figure*}[t]
   \centering
  \includegraphics[height=17cm,angle=90]{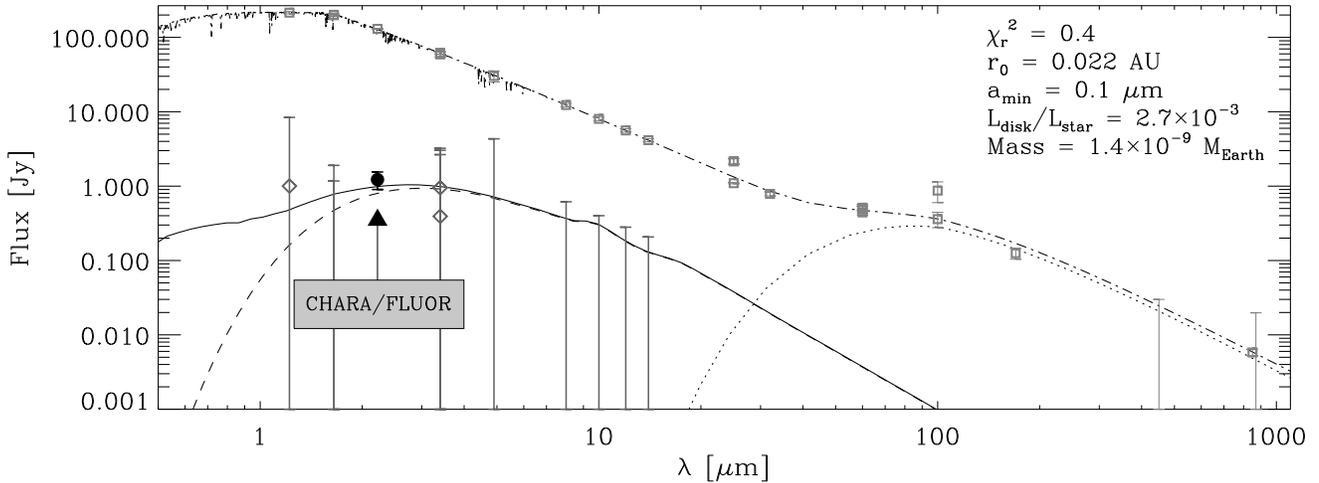}
\caption{SED of \tcets. {\it Upper plot:} Published photometric measurements are
represented by grey squares with $1\sigma$ uncertainties. A NextGen spectrum (dashed-dotted line)
is also plotted. The dotted line corresponds to a modified
$60$\,K--blackbody spectrum, highlighting the known long-wavelength excesses due to the
cold debris disk. {\it Lower plot:} residual excess in the 1--14\micron~window (diamonds,
with $1\sigma$ error bars), obtained after subtracting the best-fit stellar spectrum.
The CHARA \Kprime--band direct measurement is shown as a black dot. The solid line represents the
total emission of the inner disk of hot grains described in Sect.~\ref{sec_disk_simu}, while
the dashed line shows the thermal emission contribution of the disk. In this model,
scattered light represents about $20\%$ of the total disk emission in the \Kprime--band.}
         \label{Fig_SED}
   \end{figure*}

We compiled published flux measurements from the visible to the sub-mm, resulting in the spectral
energy distribution (SED) shown in Fig.~\ref{Fig_SED}. The stellar atmosphere SED is modeled
with a NextGen spectrum \citep{hau99} with \teff$=5400$\,K and $\log{g} = 4.5$, and scaled to match
the observed $V$-band magnitude (dashed-dotted line on Fig.~\ref{Fig_SED}). The well-known far-IR
excess beyond 60\micron~ is also shown in Fig.~\ref{Fig_SED}, where it has been fitted, for
the sake of comparison, with a modified $60$\,K--blackbody following the \citet{gre04} prescription
(dotted line). This long-wavelength emission component is associated with a reservoir of cold
material comparable in size to the solar Kuiper belt, and imaged by \citet{gre04} with
SCUBA. These cold grains orbit far beyond the 3\,AU region probed with CHARA, and we will thus
concentrate only on the warm dust content close to the star.

The 1--14\micron~SED of the assumed warm dust population 
(plotted below the star SED on Fig.~\ref{Fig_SED}) is obtained by 
subtracting the modeled photospheric emission from the measurements 
collected in the literature, except for the \Kprime--band where 
we used the excess directly measured by FLUOR. 
As in \citet{abs06}, theoretical inner disk SEDs are calculated using the optically thin disk 
model of \citet{aug99} for broad ranges of minimum grain radii 
($a_{\rm min}$), dust chemical compositions, disk inner rim positions ($r_0$) 
and surface density power law indexes.
Compared to the Vega case it appears that the fits are less constrained,
in particular because we are lacking accurate 10\micron~ observations, 
but qualitatively the model requirements are very similar. 
We have therefore adopted
the grain composition ($50\%$ glassy olivine, $50\%$ amorphous
carbons), the size distribution ($\d n(a) \propto a^{-3.7} \d a$), and the surface density profile
($\Sigma(r)\propto r^{-4}$) found to best fit the Vega inner disk. 
The disk mass within the FOV $M_{\rm dust}$ is then
obtained by a least-square fit of the photometric constraints.
The strongest constraints to the fit are the \Kprime--band FLUOR
measurement and the Spitzer/IRS observations \citep[][and Spitzer archives data]{che06}.
According to \citet{che06}, the 10--35\micron~IRS spectrum is consistent
with purely photospheric emission within an the uncertainty of ($5$\%). The non-detection
of a disk at mid-IR wavelengths is here represented as upper limits to the
disk emission at four representative wavelengths ($8$, $10$, $12$ and
$14\,\mu$m).

The best $\chi^2$ values are obtained for minimum grain sizes $a_{\rm min}$ smaller than about
$1\,\mu$m. Assuming $a_{\rm min}=0.1$\micron, in order to continue the comparison to the Vega
case, we find $r_0\simeq 0.02$\,AU (i.e. 5 stellar radii), the surface density being 
effectively sharply truncated at the sublimation distance $r_{\rm sub}(a)$ for any given grain size
if $r_0 < r_{\rm sub}(a)$. The fit restricts the possible parameter space  to $r_0 \lesssim
0.1$\,AU for $a_{\rm min} < 1$\micron~ and $r_0 \lesssim 0.03$\,AU for $a_{\rm min} > 10$\micron, although with 
a lower probability. This simple modeling demonstrates that a steep distribution
(the density scales as $r^{-4}$) of hot and small grains ($a_{\rm min} \sim 0.1\,\mu$m) can
reproduce the detected \Kprime--band excess without generating much mid- or far-IR emission. Such a
tenuous inner disk would have a typical mass of the order of $10^{-9}$\Mearth~(using $a_{\rm
max} = 1.5$\,mm) and mean collisional time-scales of a few weeks at the inner disk edge (or $M_{\rm
disk} \sim 10^{-8}$\Mearth~in the less probable case of grains with $a_{\rm min} > 10\,\mu$m).

For comparison, our zodiacal cloud is estimated from its
infrared emission to have a total mass between $10^{-9}$ and $10^{-7}$\Mearth~\citep{lei96,fix02}. 
The surprising similarity in the orders of magnitude partly comes from the much steeper radial
dependence of the surface density $\Sigma(r)\propto r^{-4}$ as compared to $r^{-0.34}$ in the Solar 
system \citep[][after integration along the vertical axis]{kel98}. We note however that adopting a 
flatter surface density profile for \tcet ($\propto r^{-1}$) 
increases the computed mass within the FOV by a factor of only 5. 
The model proposed for \tcet is obviously not a unique solution,
but it does demonstrate that exozodiacal dust disks around solar-type stars might
have structures that significantly depart from that of the Solar System (see discussion in
Sect.\,\ref{sec_conclusion}). 

\subsection{New constraints on the disk around \eeri}
In Sect.~\ref{sec_results}, we placed an upper limit on a possible \Kprime--band visibility
deficit at short baselines of $1.2 \times10^{-2}$ for \eeris. Based on the same simplistic assumption for the disk 
geometry as for \tcets, this value translates directly into a
maximum fractional photometric excess $f_{\rm CSE} = 0.6\times 10^{-2}$ ($3\sigma$).
Given that the
stars are very similar in terms of fundamental parameters, we have estimated the maximum mass of
grains in the inner disk in the framework of the model computed for \tcets. According to
archive Spitzer IRS observations  \citep[program ID:  90 see also][]{mar05} \footnote{online
presentation available at: \\http://www.stsci.edu/institute/center/information/streaming/archive/
NRDD2005/workshopOverview}, it appears that no significant mid-IR excess is detectable below $\sim
15$\micron~ around \eeris, meaning that the photometric constraints for its inner disk emission
are similar to those of \tcets. Considering that the best-fit model for \tcet produces a fractional
excess of $0.8\times 10^{-2}$ at $2\mu$m, a disk mass of $10^{-9}$\Mearth~(for the steep radial
distribution) can be regarded as a conservative upper limit for hot grains around \eeris.
A tentative fit using $f_{\rm CSE} = 0.35 \pm 0.25\times 10^{-2}$ leads to a preliminary value of $M_{\rm
disk} =5 \times 10^{-10}$\Mearth~for a $r^{-4}$ density distribution and $a_{\rm min}<1$\micron~($10^{-8}$\Mearth~for $a_{\rm min} < 10$\micron~ respectively).

We note that the absence of a detectable IR excess, hence of hot grains -- if confirmed --
could be linked to the presence of a massive planet ($\sim 1.5$\,$M_{\rm Jup}$) orbiting around
\eeri on an eccentric orbit \citep[$a \sin i = 3.4 \pm 0.4$\,AU, $e=0.70\pm0.04$,][]{ben06}. This
giant planet should influence the dynamics of hot grains in the inner region of the
debris disk, where it could clear off the dust.  Numerical simulations by \citet{the02}, 
based on the early planetary parameters derived by \citet{hat00}, 
have shown that the region inside the planetary orbit is very hostile 
to planetesimal accretion and terrestrial planet formation. 
Beyond 0.9\,AU, test particles are rapidly ejected from the system 
\citep[this limit might even be decreased down to 0.6\,AU with the
larger mass and eccentricity revised by][]{ben06}. This is also consistent 
with the recent results of \citet{jon06} for the stability of
Earth-like planets in the habitable zone around \eeris. 
Inside this dynamically unstable zone, the relative orbital velocities between 
planetesimals are so large (15--40\,km/s, Th\'ebault, private communication) 
that most collisions would result in disruption rather than accretion. 
Refractory dust grains with collision velocities above a threshold of 
20\,km.s$^{-1}$ may even be partially or completely vaporised \citep{tie94}. 
The presence of an inner reservoir of large bodies to sustain the presence of
hot grains in this system seems therefore very unlikely. 

\section{Discussion}
\label{sec_discussion}
\subsection{Small grains near the star}
The small hot grains invoked to reproduce the detected emission around
\tcet are also observed in the inner part of our own planetary system.
In the case of dust located near the Sun,
grain properties that are relevant to our study
can be found in \citet{man04}, based on {\it in situ} spacecraft measurements,
remote observations at visible and IR wavelength and theoretical
modeling. Interplanetary dust particles beyond about 0.5\,AU are known to be
concentrated in the ecliptic plane, where they form the zodiacal cloud.
Their presence is also inferred down to a few solar radii by observations
of the so-called F-corona during solar eclipses. Zodiacal light observations
can trace the 1--100\micron~size particles down to 0.3\,AU.
They have revealed a flat distribution of the dust,
with a number density decreasing proportionally to about $r^{-1}$.
Inward of 0.1\,AU (20\Rsol), large changes of the grains albedo
and polarization in the F-corona suggest a change of dust size and
composition, possibly linked to the sublimation of various dust species.
Temporary detections of an IR brightness feature during eclipses
can also be linked to a significant amount of thermal emission,
dominated by efficient absorbers like carbon grains.
These past detections have opened up a debate about the
possible existence of a dust ring composed of particles of sizes
$a < 10$\,$\mu$m, with a local density enhancement of
a factor 4 in the 1--4\Rsol~region \citep{kim98}.
There are therefore similarities with these models and the grains required to model 
the inner disk around \tcets, however, the detected emission around this latter 
appears much more prominent than in the solar system. 

\subsection{Origin of the dust}
Dust supply in the solar zodiacal cloud is supposed to be sustained by 
asteroid collisions and evaporation of short-period comets, although the
respective contributions are unknown. Mechanisms invoked to produce 
the warm dust ($T_{\rm gr} \gtrsim 300$\,K) observed in the handful of 
presently detected systems are comprised of: 
a steady-state evolution of a massive asteroid belt undergoing collisions 
and eventually sublimation, the sublimation of a ''supercomet'' captured in 
a close-in orbit \citep{bei05}, the recent collision between large planetesimals, 
and the sublimation of a swarm of comets/planetesimals possibly
triggered by a late heavy bombardment (LHB)-like event. 
\citet{wya07} have investigated these various scenarios for the few stars known 
from Spitzer spectroscopic surveys to harbour such warm grains. These are
detected through their $10$\micron~silicate feature, 
but for the vast majority they are not yet not spatially resolved. They conclude that recent and 
transient events that would have caused a population of planetesimals 
to be scattered in the inner disk from an outer belt are the most likely explanation. 
Since an LHB is believed to be triggered when migrating giant planets 
cross their mutual mean motion resonance 
\citep[1:2 resonance in the case of the solar system, ][]{gom05}, 
this could in principle occur at any time in the system's history, 
depending on the initial conditions for the planets relative positions 
and the planetesimal disk. With an age of about 10\,Gyr and a known reservoir of
potential comets (one or two orders of magnitudes more massive than our KB), 
\tcet is likely undergoing a transient event of dust production, 
whose nature remains to be determined. Precious hints for the origin 
of the dust and its production mechanism might come from future
detections of long-period giant planets or the detection of silicate features.

\subsection{Prospect for DARWIN target selection}
The bright nature of exozodiacal clouds is a potential problem for 
the direct detection of terrestrial planets around nearby stars. 
Future missions like DARWIN or TPF-I/-C, which aim to detect Earth-like planets 
around mature solar type stars, will be based on nulling interferometry 
or chronography. The presence of small grains in exozodiacal clouds 
could be a dominant noise source in both methods, depending not only on the level 
of the emission but also on its spatial distribution 
(radial extension, disk inclination, presence of structures,...). 
Earth-like planet detection in the presence of dust clouds 10~times brighter than our own zodiacal cloud 
might be very difficult, a level still two orders of magnitude below 
the current sensitivity limit of Spitzer around 10\micron.
Studying the emission of inner hot dust disks in the near- and mid-IR 
with an increased sensitivity is therefore a crucial pre-requisite 
to the selection of suitable targets for direct planet detection. 
In the case of \tcets, further investigations of 
its inner debris disk are necessary to characterise its spatial extension 
in the terrestrial planets zone and to compare its mid-IR brightness 
to that of the Solar system. Correlating the presence of such near-IR emission 
with excesses at longer wavelengths could potentially provide valuable 
diagnostics for the target selection process of these future missions.
The extension of our program to a broader sample of nearby 
Solar-type stars, either with the CHARA/FLUOR facility or with the VLTI 
instrumentation, should make it possible to carry out such analysis in the next few years.

\section{Conclusion}
\label{sec_conclusion}
We have used the short and long baselines of the CHARA Array
and the FLUOR interferometric instrument to probe the close environment of
two sun-like stars \eeri and \tcets. Our observations put a stringent upper limit on the
$2$\micron~emission of possible hot dust grains around \eeri and we derive a more precise
photospheric diameter estimation compared to our previous VLTI/VINCI results. Around \tcets, we
detect at small short baseline a visibility deficit, from which we derive a faint photometric
excess originating from the circumstellar environment. We showed in Sect.~\ref{sec_binary} that
the case of a bound or background companion can be excluded with a high probability, while a simple
distribution of warm dust grains could be consistent with the photometric constraints from
the near- to mid-IR. Preliminary modeling results show that a grain distribution with micronic or
sub-micronic minimum sizes, and with surface density peaking at a few stellar radii, is best able to
reproduce the detected near-IR excess together with the lack of significant 10--15\micron~emission (Spitzer constraint). The $2$\micron~excess detection calls for more efforts to probe the
possible $10$\micron~emission of the suspected grains around \tcet either with MIDI at the VLTI or 
with the nulling mode of the Keck interferometer. Characterising the spatial extension of 
inner exozodiacal disk emissions is a crucial precursor science on the road to future 
terrestrial planet imaging missions like DARWIN or TPF. %
\begin{acknowledgements}
We thank P.~J.~Goldfinger and Ch.~Farrington for their assistance with operation of CHARA. The CHARA Array is operated by the Center for High Angular Resolution Astronomy with support from Georgia State University and the National Science Foundation, the W.~M.~Keck Foundation and the David and Lucile Packard Foundation. We are grateful to Ph.\,Th\'ebault for fruitful discussions, that contributed to improve our analysis and our understanding of the dynamical environment of \eeri. We gratefully thank the anonymous referee for valuable suggestions and comments. O.A.~acknowledges the financial support of the Belgian National Fund for Scientific Research (FNRS) while at IAGL and of a Marie Curie Intra-European Fellowshipwhile at LAOG. This research made use of NASA's Astrophysics Data System and of the SIMBAD database, operated at CDS (Strasbourg, France). This publication makes use of data products from the Two Micron All Sky Survey, which is a joint project of the University of Massachusetts and the Infrared Processing and Analysis Center/California Institute of Technology, funded by the National Aeronautics and Space Administration and the National Science Foundation.
\end{acknowledgements}

\end{document}